\documentclass[useamsmath,useAMS,usenatbib,usegraphicx,usedcolumn]{mn2e}
\usepackage{pdflscape}
\title[Masses and luminosities of maxBCG  clusters]{A derivation of masses and total luminosities of galaxy groups and clusters in the maxBCG catalogue.}
\author[Proctor, Mendes de Oliveira, Azanha, Dupke, Overzier]
{Robert N. Proctor$^{1}$, Claudia Mendes de Oliveira$^{2}$, Luiz Azanha$^{2}$, Renato Dupke$^{1,3}$,\\
\\
\LARGE \rm{Roderik Overzier}$^1$\\
\\
$^1$  Observat\'{o}rio Nacional, Rua Gal. Jos\'{e} Cristino, 20921-400, Rio de Janeiro, Brazil\\
$^2$  Universidade de S\~{a}o Paulo, IAG, Rua do Mat\~{a}o, 1226, S\~{a}o Paulo, 05508-090, Brasil\\
$^3$  University of Michigan, Ann Arbor MI 48109, USA; Eureka Scientific Inc., Oakland CA 94602-3017, USA\\
email: rnp059@gmail.com}

\begin{document}

\maketitle

\label{firstpage}

\begin{abstract}
We report the results of a multi-waveband analysis of the masses and
luminosities of $\sim$600 galaxy groups and clusters identified in the
maxBCG catalogue. These data are intended to form the basis of future
work on the formation of the ``$m_{12}$ gap" in galaxy groups and
clusters.  We use SDSS spectroscopy and $g$, $r$ and $i$ band
photometry to estimate galaxy group/cluster virial radii, masses and
total luminosities. In order to establish the robustness of our
results, we compare them with literature studies that utilize a
variety of mass determinations techniques (dynamical, X-ray, weak
lensing) and total luminosities estimated in the $B$, $r$, $i$, and
$K$ wavebands. We also compare our results to predictions derived from
the Millennium Simulation. We find that, once selection effects are
properly accounted for, excellent agreement exists between our results
and the literature with the exception of a single observational
study. We also find that the Millennium Simulation does an excellent
job of predicting the effects of our selection criteria. Our results
show that, over the mass range $\sim10^{13}-10^{15}$ M$_{\odot}$,
variations in the slope of the mass-luminosity scaling relation with
mass detected in this and many other literature studies is in part the
result of selection effects. We show that this can have serious
ramifications on attempts to determine how the mass-to-light ratio of
galaxy groups and cluster varies with mass.

\end{abstract}


\section{Introduction}

In 1933 Fritz Zwicky applied the virial theorem to a galaxy cluster
(the Coma cluster) for the first time \citep{Zw33}. He concluded that,
in order to explain the cluster's dynamics, the average density of the
cluster had to be several hundred times greater than that indicated by
estimates of the mass of luminous material observed in the cluster's galaxies
alone. Confirmed later in the Virgo cluster \citep{Smith36}, these
were, of course, the first direct observations indicating the presence
of dark matter (Zwicky's ``Dunkle Materie").

Zwicky's methodology of measuring the mass of clusters and comparing
this mass to the total light emitted by the galaxies in the cluster is
still in use today. Indeed, the measurement of mass-to-light ratios
in galaxy groups and clusters has become commonplace since these
discoveries, as knowledge of how the mass of baryonic and non-baryonic 
matter are distributed within clusters provide important clues as to how 
they, and the galaxies within them, were formed. 

However, the techniques and data involved in measuring both the masses and
luminosities of clusters have become significantly more sophisticated
since Zwicky's original discovery.  For instance, with the coming of
large telescopes and large scale surveys, the measurement of
group/cluster luminosities has advanced, permitting direct observation
of galaxies further down the luminosity function and to higher
redshifts, while mass estimates can now be based upon cluster dynamics
(as performed by Zwicky and Smith), or in more recent advances, X-ray
properties of cluster halos or gravitational lensing. However, due to
the variations in methodologies adopted in works in the literature,
the received wisdom has become that it is difficult to make direct
comparisons between studies.

In this work we examine this issue, comparing the results of our
analysis of the masses and luminosities of $\sim$600 galaxy groups 
and clusters identified in the
maxBCG catalogue with the results of the Millennium Simulation and
five other recent studies in the literature that use a variety of
methodologies and data sources. Our analysis is performed in four
wavebands ($B$, $r$, $i$ and $K$) and includes systems of mass ranging from
$\sim$10$^{13}$M$_{\odot}$ to 10$^{15}$M$_{\odot}$. We examine both
the extensive areas of agreement, as well as areas of disagreement,
between these studies and identify the issues that must be considered
before making comparisons between studies.

In future papers we will use the data presented here to perform an
analysis aimed at identifying the group/cluster properties driving
differences in the ``$m_{12}$ gap" (the luminosity difference between
the brightest and second brightest galaxies within the central regions
of groups and clusters).

Throughout this paper both our data and literature data are presented
assuming a H$_0$=70 km s$^{-1}$ Mpc$^{-1}$, $\Omega_{\Lambda}$=0.7,
$\Omega_M$=0.3 cosmology. For expressing luminosities in solar units
we used the values M$_{r,\odot}$=4.67 mag, M$_{i,\odot}$=4.48 mag,
M$_{B,\odot}$=5.33 mag, M$_{K,\odot}$=3.28 mag.

\section{Data}
Our sample was selected from groups identified in the maxBCG catalogue
of \citet{Koester07b}. This catalogue was constructed from the SDSS
photometric survey using a cluster finding algorithm based on three
well defined properties of galaxies in clusters: spatial clustering,
the presence of a ``red-sequence" and the presence of a BCG
(brightest cluster galaxy) at the center of the cluster
\citep{Koester07a}. Using this method \citet{Koester07b} identify more
than 13,000 clusters.  Using DR9 of the SDSS-III \citep{Ahn12}, we
then selected all the clusters from the maxBCG catalogue that have a
BCG with a spectroscopic redshift in the redshift interval $z=0.05$ to
$z=0.16$ and possessing 6 or more spectroscopically confirmed
members (as described in Section \ref{select}). 

The upper end of the selected redshift range was chosen in order to
ensure a substantive sample at all values of the ``$m_{12}$ gap", as
it is necessary to probe out to redshifts of $\sim$0.15 in order to
obtain a reasonably large sample of the rare ``fossil groups" (defined
as systems with $m_{12}>2$). The lower end of the redshift range was
chosen to minimize the impact of the redshift dependent selection
criteria of the SDSS spectroscopic survey which only sampled galaxies
with apparent magnitude brighter than 17.7 mag in the $r$ band, and to
ensure that the peculiar velocities of the groups and clusters studied
have minimal impact on distance determinations and their associated
distance moduli, and thus ensuring accurate luminosity estimates. A
minimum of 6 spectroscopically confirmed members was required as the
mass estimates made in this work are based on velocity dispersion
estimates whose errors are large when the number of spectroscopically
confirmed members is low. Indeed, even with a sample of 6
spectroscopically confirmed members, errors in log(mass) estimates are
of order of 0.4 dex (or a factor of 2.5 in mass).

\subsection{Data from the literature}
In this work we compare our results to those of five recent studies.
These studies are: \citet{Gir02}, \citet{Eke04b}, \citet{Ram04},
\citet{Pop07} and \citet{Sheld09}. These studies span four wavebands
($B$, $r$, $i$ and $K$). In each case the results of these studies
were, if necessary, converted to the H$_0$=70 km s$^{-1}$ Mpc$^{-1}$,
$\Omega_{\Lambda}$=0.7, $\Omega_M$=0.3 cosmology used throughout this
paper. Where necessary, small corrections were also made to ensure
that the same solar luminosities were used in the estimation of
group/cluster luminosities and that the data are k-corrected to the
same redshift ($z=0$). We note that the literature studies above are
generally based on groups/clusters at lower redshift than our data,
and a minimum of 10 spectroscopically confirmed members is generally
applied (rather than the 6 used here). The studies using dynamical
mass estimates therefore generally possess average mass errors that
are smaller than ours by, in the worst case, $\sim$0.1 dex.

\subsection{The Millennium Simulation}
\label{MS}
To supplement our analysis and provide a means of testing the impact
on observations of errors and selection effects, we also use the
galaxy recession velocities (for dynamical mass estimation) and 
luminosity data from the Millennium Run
dark matter simulation in the WMAP1 cosmology (H$_0$=73 km s$^{-1}$
Mpc$^{-1}$, $\Omega_{\Lambda}$=0.75, $\Omega_M$=0.25) from Springel et
al. (2005), together with the semi-analytic galaxy modeling technique
as described in Guo et al. (2011). The semi-analytic galaxy catalogs
were converted into an observer's light-cone geometry matched to the
SDSS survey by \citet{Hen12} using the models of \citet{BC03},
assuming a \citet{Chab03} initial mass function. \citet{Hen12}
estimated total luminosities in a broad range of photometric bands
from the B band to the infrared. In this work we use the r band data
only. These data were converted to the H$_0$=70 km s$^{-1}$ Mpc$^{-1}$ 
cosmology used throughout this paper. The corrections for
$\Omega_{\Lambda}$ and $\Omega_M$ were deemed insignificant for this
work as they cause differences of only $\sim$0.01 mag and $\sim$0.4\%
in the distance moduli and angular sizes, respectively. Data for all
galaxies brighter than 22 mag in the r band (the photometric limit of
the SDSS data used in our analysis) in simulated halos at redshifts
less than 0.16 were obtained, and the positions and $r$ band
luminosities of each galaxy, along with the $M_{200}$ of the parent
halos, were tabulated.

\subsection{Spectroscopic selection and analysis}
\label{select}
The identification of spectroscopically confirmed cluster members was
performed by an iterative procedure. First, each cluster was assumed
to possess a velocity dispersion ($\sigma$) of 600 km
s$^{-1}$. $R_{200}$ was then estimated assuming the relation of
\citet{Carl97};

\begin{equation}
R_{200} = \frac{\sqrt{3}\sigma}{10H(z)}~\rm{Mpc}. 
\label{R200}
\end{equation}

We then searched the SDSS spectroscopic catalogue for all galaxies
within the estimated $R_{200}$ that possess redshifts differing from
the BCG redshift by less than 1500 km s$^{-1}$
(i.e. $\pm2.5\sigma$). The value 2.5$\sigma$ was selected in order to
allow for the inherent scatter expected from the (almost Gaussian)
distribution of velocities observed in clusters, while at the same
time minimizing the likelihood of selecting interlopers. Once all the
members within $R_{200}$ were identified their redshifts were used to
calculate a new velocity dispersion using the \citet{Beers90}
bi-weight estimator for systems with more than 10 members, and the
\citet{Beers90} gapper method for systems with 10 or less members.
For cases where the initial velocity dispersions were an underestimate
this resulted in an increase in the velocity dispersion (and hence
$R_{200}$) estimate, as well as increasing the permitted velocity
range ($\pm2.5\sigma_{\rm{new}}$).  Conversely, for systems in which
the initial values were an overestimate the new values were
lower. This process was iterated a number of times. We found that,
generally, convergence was achieved after only two or three
iterations. However a total of 10 iterations were performed.  This
permitted the use of the value from the last (10th) iteration as the
final result for each group/cluster, while the RMS scatter in the last
6 iterations was taken as an additional uncertainty in the velocity
dispersion and which was added in quadrature to the formal velocity
dispersion error.  We note that this additional error is zero for the
majority of systems in which the procedure converged.

At this stage, groups/clusters with less than 6 spectroscopically
confirmed members within $R_{200}$ were discarded from the
sample. Clusters for which the average recession velocity of all
the galaxies \emph{except} the BCG differed from the recession
velocity of the BCG by more than 400 km s$^{-1}$ were also excluded,
as these have a high probability of being either highly
disturbed (i.e. non-virialized) systems or false/contaminated
detections due to line-of-sight effects.

\subsection{Mass estimation} 
\label{massest}
The masses of the remaining maxBCG clusters were estimated using:

\begin{equation}
M_{200} = \frac{3}{G}\sigma_{200}^2R_{200}.
\label{mass}
\end{equation}

This equation can be reformulated using Equation \ref{R200} to give a
more convenient form:

\begin{equation}
\rm{log}(M_{200}) = 3\rm{log}(\sigma_{200})+6.24-\rm{log}(E(z)),
\label{logmass}
\end{equation}

\noindent where, for a flat Universe:

\begin{equation}
E(z)=\sqrt{\Omega_M(1+z)^3+\Omega_{\Lambda}}.
\label{Ez}
\end{equation}

However, it is helpful to note that log(E(z)) is small ($\sim$0.02 at
the $z=0.1$ median of the data presented here) and varies by only
$\pm$0.02 over the whole redshift range of the clusters reported
here. The resultant masses spanned the range from $\sim$10$^{13}$ to
$\sim$10$^{15}$ M$_{\odot}$.\\

We note that four of the five literature studies to which we compare
our results \citep{Gir02, Eke04b, Ram04, Pop07} use Equation
\ref{mass} to estimate masses. However, both \citet{Gir02} and
\citet{Pop07} make corrections for a surface pressure term. A
comparison of the masses presented in \citet{Gir02} with masses
estimated using their velocity dispersions and Equation \ref{mass}
yields an offset of 0.00 dex with RMS scatter of only 0.12 dex. This
demonstrates that this term is small and of no significance to the
results presented here.  \citet{Eke04b}, on the other hand, use the
spatial distribution of the galaxies in each system to estimate
$R_{200}$ rather than Equation \ref{R200}, basing their values of
radius on a calibration against cosmological simulations
\citep{Eke04a}\footnote{N.B. We have converted the b$_j$ given in
  \citet{Eke04b} to the $B$ band using the relation $L_{bj}=1.1L_B$
  given by their Equation 4.3}. Finally, \citep{Sheld09} utilize SDSS i band
photometry and masses derived from weak gravitational lensing.
 
As a first test of our dynamical mass estimates, we performed a search
of the literature for groups/clusters in our study for which masses
derived from X-ray properties have either been reported or can be
estimated. A total of 24 groups from our study were found to have
masses based on a variety of X-ray properties reported in the
literature
\citep{Donahue05,Jia06,Lemze08,Zibetti09,Akamatsu11,Pea11,Miller12,Owers14}.
For the majority of of these studies
\citep{Donahue05,Zibetti09,Akamatsu11,Pea11,Miller12,Owers14} X-ray
masses were estimated from X-ray temperature using the the Tier 1
relation from \citet{Sun09} (their Table 6). However, the X-ray masses
taken from \citet{Jia06} were deived using X-ray temperature and
electron density, while in the case of \citet{Lemze08}, weak lensing
and the X-ray luminosity profile were employed to determine cluster
masses.  The masses derived from all the above studies assume
hydrostatic equilibrium.

\begin{figure}
\centering
    \includegraphics[width=0.4\textwidth,angle=-90]{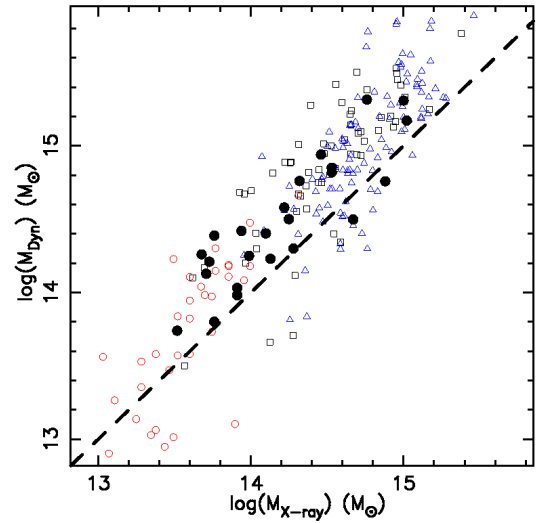}
    \caption{Comparison of dynamical mass, estimated as detailed in
      the text, with masses derived from X-ray data. Black dots are
      clusters in our study (whose dynamical masses are estimated
      here) with X-ray masses from the literature.  Open squares are
      derived from the velocity dispersion and X-ray gas mass data
      from \citet{Zhang11} (see text). Open circles and triangles are
      derived from the velocity dispersion and X-ray temperature data
      of \citet{OP04} and \citet{Wu99}, respectively. The dashed line
      shows the one-to-one locus. At masses greater than $\sim
      10^{14}$ M$_{\odot}$ there is a clear, constant offset between
      X-ray and dynamical masses, while below this mass there is a
      suggestion of a change in slope in the correlation.}

\label{xray}
\end{figure}

The results of the comparison of dynamical masses to those derived
from X-ray data are shown as solid points in Fig. \ref{xray}. Also
shown in this figure as open squares are total cluster masses derived
from \citet{Zhang11} X-ray gas masses assuming the relation derived by
\citet{Mahdavi13} between X-ray gas masses and total cluster masses.
This relation was calibrated by comparing gas masses from X-ray
analyses to the results of total cluster masses from weak
lensing. Dynamical masses for these data were derived using the
velocity dispersions provided in \citet{Zhang11} and Equation
\ref{logmass}. Velocity dispersions and masses derived from X-ray
temperature were also taken from \citet{OP04} and \citet{Wu99}. For
these data, dynamical masses were derived using Equation
\ref{logmass}, while X-ray masses were derived from the X-ray
temperatures using \citet{Sun09} as above. The results of
\citet{OP04} and \citet{Wu99} are shown in Fig. \ref{xray} as open
circles and triangles respectively. Again, hydrostatic equilibrium is
assumed in all cases, and all the above results are for masses within
$R_{200}$.  Where necessary, we assumed $R_{200}$=1.54$R_{500}$,
consistent with the values of $R_{200}$ and $R_{500}$ for clusters
given in \citet{Ettori10}, yielding $M_{200}$=1.46$M_{500}$.

Now, given the inhomogeneity of these data, and the need to convert
between $M_{500}$ and $M_{200}$, it would be unwise to over-interpret
this plot. However, we note that in Fig. \ref{xray}, systems with
masses above $\sim$10$^{14}$ M$_{\odot}$, although running parallel to
the one-to-one line, the X-ray masses appear systematically offset
from the dynamical masses by $\sim$0.15 dex. On the other hand, the
systems with masses lower than $\sim$10$^{14}$M$_{\odot}$ (which are
from \citet{OP04}), the data straddle the one-to-one line, but appear
to exhibit a slope greater than one. This change in gradient at low
masses was noted in \citet{OP04} when they considered their
$L_X-\sigma$ and $\sigma-T_X$ plots (their Sections 7.2 and 7.3) and
was also noted in \citet{Helsdon00}. A break in the slope of the X-ray
scaling relations at a temperature of about 1 keV (corresponding to a
mass of $\sim$10$^{14}$ M$_{\odot}$) is also evident in a number of
studies (e.g. \citet{Xue00} and \citet{Harrison12}). However, it has
been disputed in other works, e.g. \citet{Mulchaey00}. Therefore,
there remains no real consensus as to whether the apparent change
in gradient in the $L_X-\sigma$ and $\sigma-T_X$ scaling relations
represents a true physical effect or is rather an artifact of large
errors and selection effects in low mass systems (see \citet{OP04} for
a more detailed description of the controversy).

It is also interesting to perform a similar comparison using the
Millennium Simulation data, in this case comparing dynamical mass to
the M$_{200}$ values given in the database. The masses presented in
the database were calculated by simply adding the masses of all the
particles within the region of each halo where the average density is
200 times the critical density. In order to calculate dynamical masses
we applied the same analysis as that applied to our observational
data. i.e. assuming that the brightest cluster galaxy lay at the
center of the cluster and measuring recession velocities with respect
to the central galaxy using the \citet{Beers90} bi-weight and gapper
estimators as described in Section \ref{select} to estimate velocity
dispersion.  Dynamical masses are then estimated using Equation
\ref{logmass}. This analysis was performed under the assumption of a
variety of selection criteria, including those describing our
observational data, which we recall were selected to have redshifts
between 0.05 and 0.16, and have 6 or more members with apparent
magnitudes brighter than the 17.7 mag within $R_{200}$. For
each analysis, the median of the dynamical mass data in bins of
Millennium Simulation mass were calculated. We note that our use of the
median in these analyses ensures that the effects of the asymmetrical
errors in dynamical mass are essentially eliminated.

\begin{figure}
\centering
    \includegraphics[width=0.4\textwidth,angle=-90]{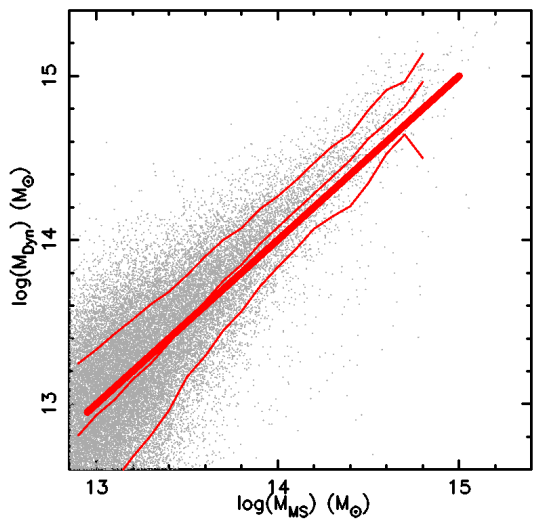}
    \caption{Comparison of masses using our dynamical mass estimation
      method (see text) with masses from the Millennium
      Simulation. The thick red line shows the one-to-one locus, while
      the thin lines show the median and 15/85 percentiles of the
      dynamical masses in bins of 0.1 dex in Millennium Simulation
      mass. As well as an offset at high masses, the data exhibit and
      increase in slope at masses below 10$^{14}$ M$_{\odot}$.}
\label{MRmasscomp}
\end{figure}

The comparison between Millennium Simulation masses and dynamical
masses is shown in Fig. \ref{MRmasscomp}. In this plot the dynamical
masses are those for all systems with a minimum of 6 members, but with
no redshift or member luminosity selection limits applied. Reasonable
agreement is found.  At masses greater than $\sim$10$^{14}$
M$_{\odot}$ the Millennium Simulation masses show an offset from the
median dynamical masses of $\sim$0.1 dex, similar to the trend in the
X-ray data in Fig. \ref{xray}. Below $\sim$10$^{14}$ M$_{\odot}$, the
correlation between M$_{MS}$ and M$_{Dyn}$ steepens, with the gradient
increasing from 1.05 above $\sim$10$^{14}$ M$_{\odot}$ to 1.17 below
this mass. Although statistically only marginally ($\sim$2$\sigma$)
significant, this steepening at low masses is again at least
qualitatively similar to the X-ray data.  Some care must be taken in
the interpretation of these differences, since given the large
uncertainties in the semi-analytic models that describe the baryonic
physics, there is no guarantee that the Millennium Simulation data
match the real Universe in this regard.  However, it is interesting to
note that offset at high masses and the increased gradient at lower
masses are at least qualitatively consistent with the trends in the
X-ray data. This therefore raises the possibility that this represents
a real systematic bias in dynamical masses, perhaps due to a bias in
the distribution of galaxies with respect to the underlying dark
matter halo or perhaps representing a breakdown in the assumptions
made in deriving dynamical masses (i.e. the assumption of full
virialization and/or that the systems are well described as isothermal
spheres).

\subsection{Photometric analysis}
Having established estimates of the masses of the clusters in our
sample we must next estimate the total luminosity of each of the
systems.  Therefore, for each of the clusters selected above, the $g$, $r$
and $i$ band photometry of all galaxies within 4$R_{200}$ of the BCG
were obtained from the SDSS photometric survey. We use the SDSS model magnitudes which were estimated using a combination of De Vaucouleurs and exponential profiles. Galaxies with SDSS
spectroscopic redshifts were k-corrected to zero redshift using the
colour dependent functions of \citet{Chil10}.

For each cluster, and in each waveband ($g$, $r$ and $i$), the total
luminosity of the galaxies within $R_{200}$ was then estimated in a
two step process. First, the luminosities of all the spectroscopically
confirmed galaxies within $R_{200}$ of each group were co-added to
derive the total luminosity. Next, the total luminosity of all
galaxies in the photometric survey with apparent magnitudes fainter
than 17.7 mag in the $r$ band (the lower limit of the spectroscopic
survey) and brighter that 22.2 mag (the 95\% completeness limit of the
photometric survey) were co-added. We note that modeling of the
\citet{Blan03} SDSS r band luminosity function indicates that even
in our most distant clusters the contribution to the total light of
galaxies fainter than our upper limit of 22.2 would be less than 2.5\%
(i.e a log(luminosity)=0.01). This contribution is therefore deemed
insignificant to our results, and is ignored.

A colour cut excluding all galaxies 0.2 mag redder in $(g-r)$ than the BCG
was applied in order to minimize the impact of contamination by high
redshift interlopers. In order to estimate the background
contamination, the same process was applied to seven concentric annuli
between 3$R_{200}$ and 4$R_{200}$. These annuli were selected to have
an area equal to the area of the central region within R$_{200}$.  In
each of the $g$, $r$ and $i$ bands the median value of the seven annuli was
then taken as the background level and this value was subtracted from
the value calculated within $R_{200}$ yielding an estimate of the
total luminosity of faint galaxies within $R_{200}$. These
luminosities were also then k-corrected to zero redshift using the
functions of \citet{Chil10}. The total luminosity in faint galaxies
was then added to the total luminosity in spectroscopically confirmed
galaxies to yield a total luminosity for each group.

However, in some cases the background luminosity in faint galaxies
exceeded the estimated total luminosity of \emph{all} the galaxies in
the cluster, resulting in negative values for the luminosity of the
cluster, clearly indicating that the background level was
overestimated. Such clusters were excluded from our analysis. In other
cases, despite high background levels, the total luminosities of the
clusters remained positive raising the possibility that the clusters
had, indeed, very low luminosity compared to the background against
which they are projected. In such cases we excluded only clusters in
which the background level exceeded five times the luminosity of faint
galaxies in the cluster, deeming the uncertainties in the background
level to be too large for a reliable total luminosity estimate.\\

After all the selection cuts above were made we retained a total
of 614 clusters.\\

In order to compare literature values in the $B$ and $K$ bands we used the
colour dependent functions of \citet{Blan07} ($B$ band) and
\citet{Yaz10} ($K$ band) to estimate the luminosities of each galaxy in
each cluster in these bands. The total luminosities in the bands were
then also estimated as above.

\section{Results}
In this section we report our results and compare them to five recent
works from the literature that measure masses and luminosity ratios in
significant samples of groups/clusters in the $B$, $r$, $i$ and $K$
bands (recall that, for our results, $r$ and $i$ are taken directly
from the SDSS database while $B$ and $K$ are generated from the SDSS
data using the colour dependent transforms of \citet{Blan07} and
\citet{Yaz10} respectively). A full table of results is available
from the authors and will be made available on-line.

\begin{figure*}
\centering
    \includegraphics[width=0.75\textwidth,angle=-90]{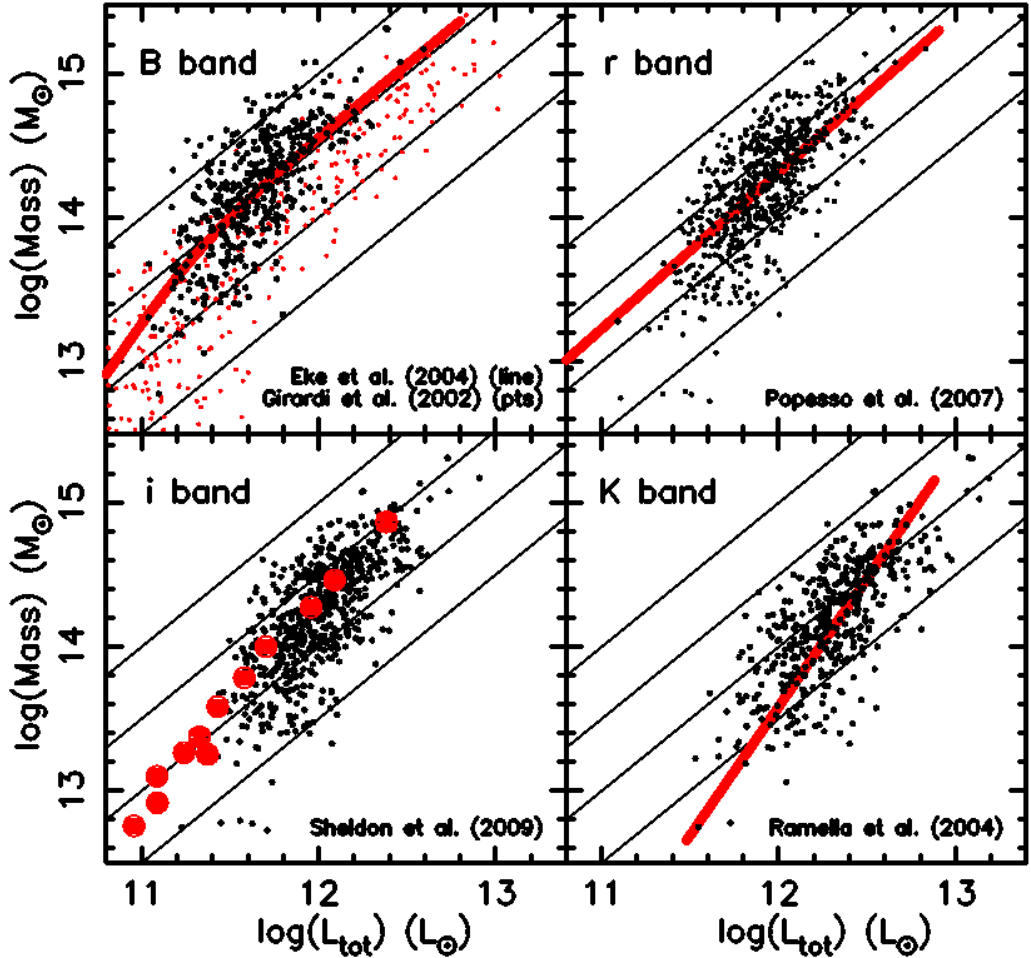}
    \caption{Mass is plotted against total group/cluster luminosity. 
      Black dots are our data, while red points and lines are
      taken from literature studies (as detailed in the bottom right
      of each panel). The four black lines in each panel are lines of
      constant log of the mass-to-light ratio([M/L]), with values of
      3.0, 2.5, 2.0 and 1.5, decreasing from upper left to the lower
      right. Agreement between our data and literature data is
      generally good at high masses, but at lower masses and in the
      comparison to \citet{Gir02} in the B band discrepancies are
      evident.}
    \label{lxm}
\end{figure*}

Fig \ref{lxm} shows mass against luminosity for the clusters in our
final sample in all four bands (436, 614, 611 and 427 clusters in $B$,
$r$, $i$ and $K$ bands, respectively). Lines of constant log of the
mass-to-light ratio ($[M/L]$=log$(M/M_{\odot}$)-log$(L/L_{\odot}$)) are
shown for $[M/L]$=3.0, 2.5, 2.0 and 1.5 (decreasing to the lower
right).

In the figure our data are shown as black points, while in each panel,
data from the literature are also presented as either red points (when
actual data are available) or red lines (when the literature provide
fits to data). Note that in all cases the literature data have been
converted (when necessary) to a H$_0$=70 km s$^{-1}$ Mpc$^{-1}$,
$\Omega_{\Lambda}$=0.7 cosmology, k-corrected to $z=0$ and adjusted
the solar luminosities given in the introduction.

Fig. \ref{lxm} generally shows good agreement between our data and the
literature studies shown in red, particularly at masses higher than
$>$10$^{14}$ M$_{\odot}$. However, there appear to be two areas of
poorer agreement - the first in the comparison to the study of
\citet{Gir02}, and the other at the low mass end of our mass range
(i.e. $<$10$^{14}$ M$_{\odot}$) in the $B$, $r$ and $i$ bands.

Addressing first the poor agreement with the study of \citet{Gir02} in
the $B$ band, there is clearly an offset between our data and the
results of \citet{Gir02} at higher masses ($>$10$^{14}$ M$_{\odot}$),
with the average of log($L$) within $\pm$0.1 dex of
log($M$) =14.25 being 11.67 and 11.97 for our and \citet{Gir02} data
respectively - i.e. a 0.3 dex (or factor of 2) difference in luminosity at
a given mass, and therefore in the mass-to-light ratio. 
However, we note that our results \emph{are} consistent with the study of Eke
et al (2004b) in the $B$ band, as well as with the studies of
\citet{Pop07}, \citet{Sheld09}, \citet{Ram04} in the $r$, $i$ and $K$
bands respectively. We then conclude that the Girardi et al. (2002)
study suffers from an, as yet unidentified, systematic difference in
their luminosity estimates with respect to other recent studies. We
assume the offset to be in the luminosity direction because, as noted
in Section \ref{massest}, we have already shown that the mass
estimation methodology of \citet{Gir02} is completely consistent with
the simple dynamical mass estimates made in this paper.

Turning our attention to the disparities at low masses, our results
clearly show a pronounced increase in the slope of the mass-luminosity
relation as we go to low masses. However, we note that many of the
literature studies presented in Fig. \ref{lxm} appear to show similar,
albeit smaller, increases in the slope of the mass-luminosity relation
at the low mass end. The effect is clearly seen in the studies of
\citet{Gir02}, \citet{Eke04b} and can also be seen in the data of
\citet{Sheld09} in which the gradient decreases from $\sim$1.6 below
10$^{14}$ M$_{\odot}$ to $\sim$1.3 above this mass. The power-law fits
of of \citet{Pop07} or \citet{Ram04} do not, perforce, exhibit such
gradient changes. However, in the case of \citet{Ram04}, the study
contains very few points at the low masses of interest here, while the
\citet{Pop07} study presents data selected from two catalogues, one
optical, the other an X-ray catalogue. Examination of Fig. 9 of
\citet{Pop07} shows that the optically selected sample \emph{does}
exhibit an increase in the slope of the mass-luminosity relation at
the lowest masses, while the X-ray selected systems do not. This
suggests that this trend may be, at least in part, the result of
optical selection effects (a point that was also raised in the recent
paper of \citet{Mulroy14}). In fact, our study, which requires at least
6 bight galaxies within $R_{200}$ of galaxies in the redshift range
$z=0.05$ to $z=0.16$, is generally biased towards higher redshifts
than the other studies which are dominated by much closer
systems. This raises the possibility that this optical selection
effect may be playing a stronger role shaping the locus of our results
than in the other studies presented here. To examine this possibility
in more detail we again turn to the Millennium Simulation.

\begin{figure}
\centering
    \includegraphics[width=0.4\textwidth,angle=0]{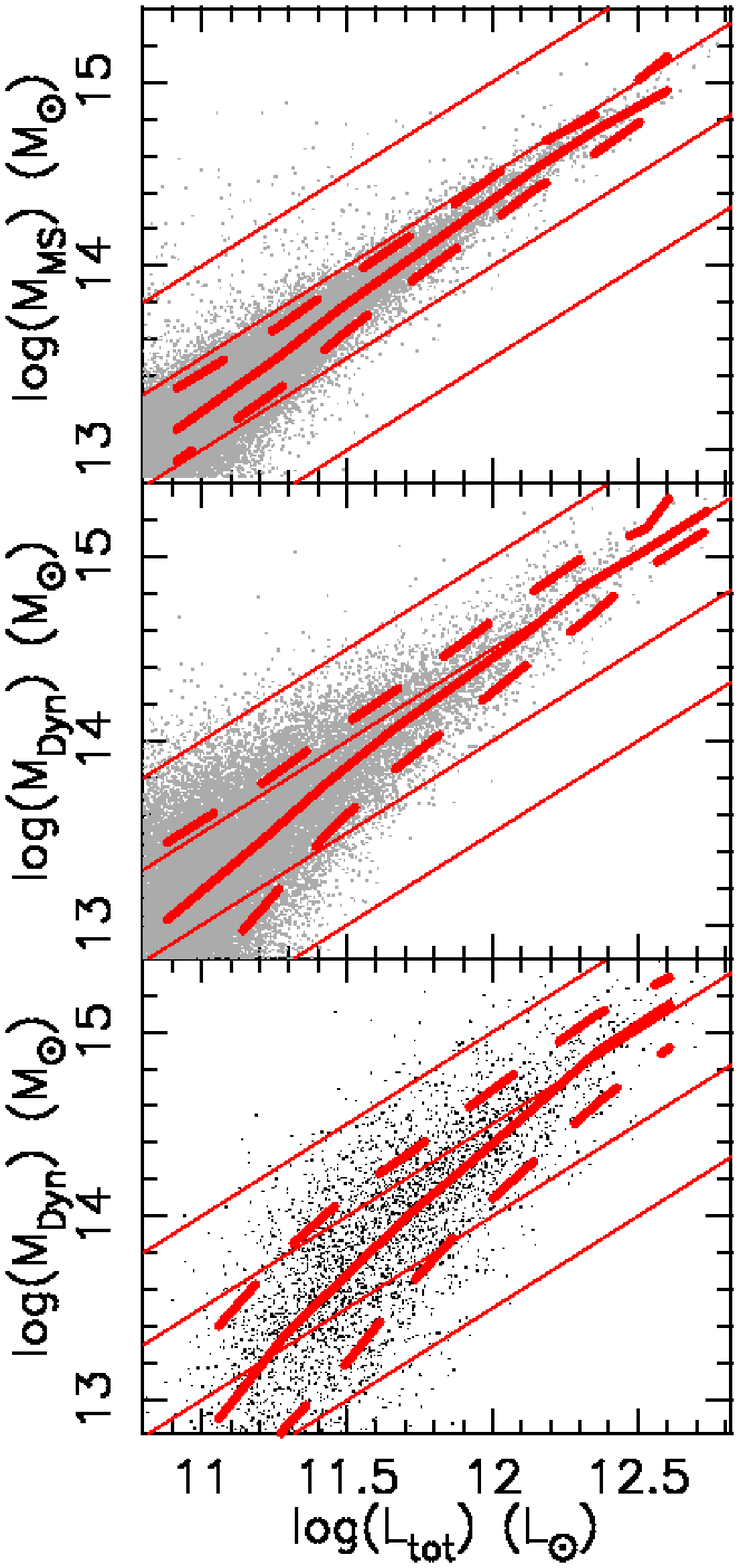}
   \caption{The r band mass-luminosity relation of groups/clusters in the
     Millennium Simulation. In the top panel Millennium Simulation
     mass (rather than dynamical mass) is plotted, while the
     bottom two panels show dynamical mass.  In the top two panels all
     systems with redshift less than 0.16 are shown, while in the
     bottom panel our observational selection criterion of a minimum of 6 bright
     galaxies within $R_{200}$ of systems at 0.05$<z<$0.16 is
     applied. Solid and dashed red lines are the median with 15th and
     85th percentiles, respectively.}
\label{mrlxm}
\end{figure}

In Fig. \ref{mrlxm} we show the mass-luminosity relation for groups
and clusters in the Millennium Simulation. The medians of the
dynamical mass data in bins of luminosity are shown as solid red
lines. The 15th and 85th percentiles are shown as dashed lines. In the
top panel we plot the masses and luminosities taken directly from the
simulation. In the bottom two panels we plot dynamical masses derived
from the simulation (see Section \ref{MS}). In the top and middle plots all
groups and clusters with 6 or more member galaxies (irrespective of
their luminosities or redshifts) are plotted. In the bottom panel we apply the
SDSS selection criteria - i.e. 6 or more galaxies brighter than
  17.7 mag in the r band, and redshifts in the range
0.05$<z<$0.16. The bottom panel therefore represents the Millennium
Simulation prediction for the locus of our observational data.

\begin{figure}
\centering
    \includegraphics[width=0.4\textwidth,angle=-90]{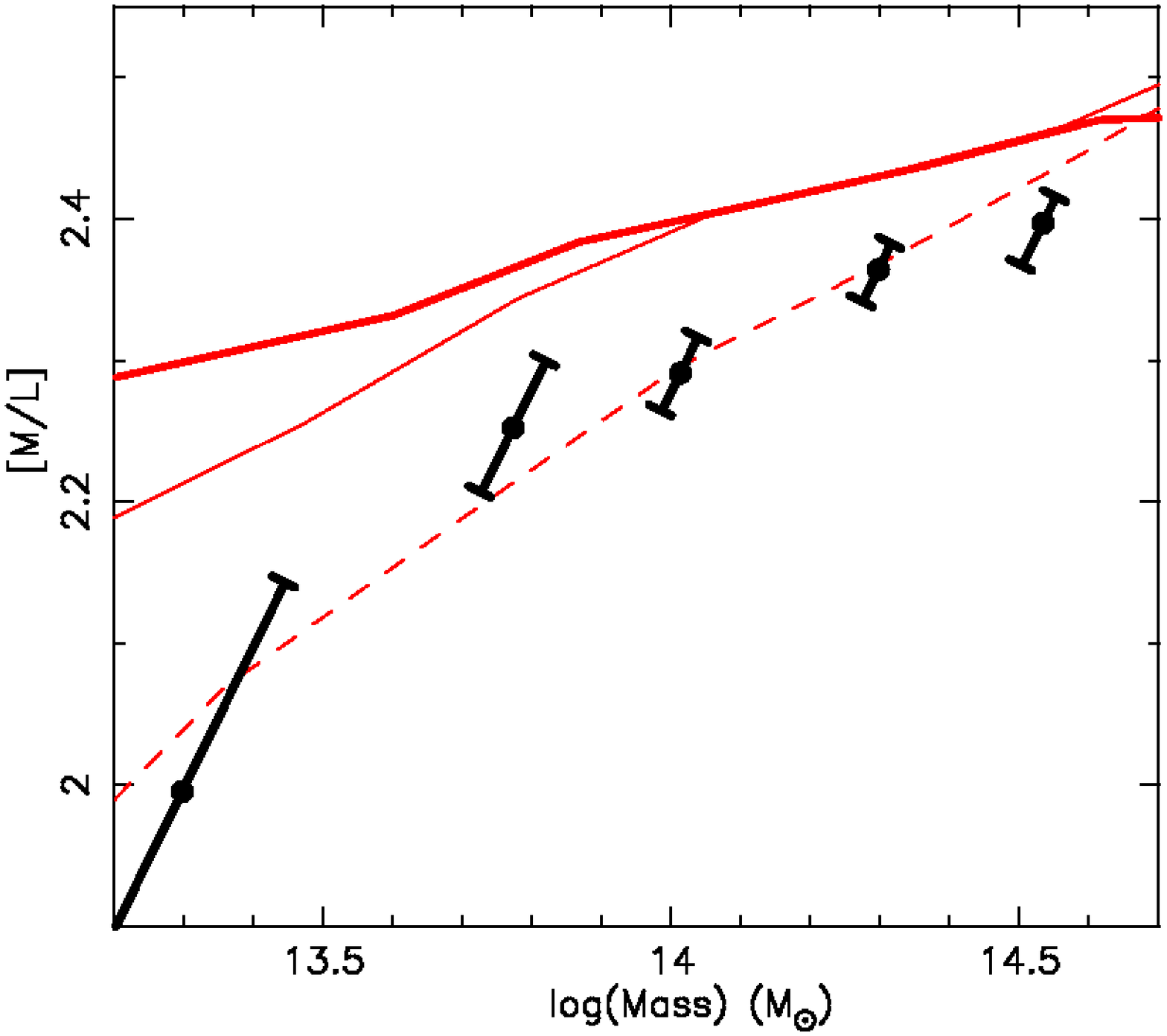}
    \caption{Mass-to-light ratio in the r band are plotted against against mass. The medians of
      observational data and their (correlated) uncertainties are
      shown as black points with error bars. Millennium Simulation data
      are shown as red lines. Data taken directly from the Millennium
      Simulation are shown a thick solid line, while the data for
      dynamical masses derived from the simulation with no selection
      criteria applied (see text) are shown a a thin red line. The
      data from the simulation using dynamical mass and with selection
      criteria applied is shown as a dashed line. The simulation
      clearly reproduces the observational data extremely well.}
\label{m2l}
\end{figure}

Comparing first the top two panels, the displacement and non-linearity
in the relation between Millennium Simulation mass and dynamical mass
identified in Fig. \ref{MRmasscomp} is also evident when comparing
these two plots. A marked increase in scatter is also clearly seen in
the dynamical mass data.

Comparison of the bottom two panels shows the effect of requiring 6 or more
bright members ($r<17.7$ mag) within $R_{200}$ and restricting the
redshift range to 0.05$<z<$0.16.  Clearly, at low masses, when
combined with the requirement of a minimum number of bright members
for dynamical analysis, this redshift restriction causes only the
brightest groups and clusters at any given mass to be selected, thus
introducing another strong non-linearity into the apparent mass-luminosity
relation. What is more, comparison of the bottom plot with the r band data in
Fig. \ref{lxm} suggests that the Millennium Simulation is reproducing
the observational data extremely well.

\begin{table}
\begin{centering}
\caption{The offsets from the median of the 15th and 85th percentiles in three masses bins in both our data and the Millennium Simulation (MS).}
\begin{tabular}{|c|c|c|c|c|}
\hline
      & \multicolumn{2}{c|}{Data}& \multicolumn{2}{c|}{MS}\\
\hline
log(M) (M$_{\odot}$)  & 15\%   &    85\%  &  15\%   &   85\% \\
\hline
13.3  & -0.52 &   0.39  &  -0.50 &   0.45\\
14.0  & -0.32 &   0.30  &  -0.39 &   0.33\\
14.7  & -0.27 &   0.27  &  -0.28 &   0.25\\
\hline
\end{tabular}
\label{spreads}
\end{centering}
\end{table}

In order to demonstrate this, in Fig. \ref{m2l} we plot the
predictions of the simulation for the mass-to-light ratio against mass
scaling relation and compare to the medians of the observational data
and their uncertainties. We choose to show mass-to-light ratio here as
the relatively small dynamical range of this parameter serves to
emphasize similarities and differences between the various data sets,
while the binning of the data has significantly reduced the correlated
errors which would otherwise make interpretation of such a plot
difficult.

In Fig. \ref{m2l} the data taken directly from the Millennium
Simulation is shown as a thick solid line. For ease of comparison, this
line has been shifted by the $\sim$0.1 dex offset between Millennium
Simulation and dynamical masses first identified in
Fig. \ref{MRmasscomp} and also evident in Fig. \ref{mrlxm}. The data
for dynamical masses with no luminosity or redshift selection criteria
applied is shown as a thin solid line, while the dynamical mass data
with luminosity and redshift selection criteria applied is shown as a
dashed line.

The changing slope in the Millennium Simulation mass against dynamical masses
relation noted in Fig. \ref{MRmasscomp} at $\sim$10$^{14}$ M$_{\odot}$
is again evident in this plot, with the two solid lines exhibiting the
same shallow slope above this mass, but steepening and diverging
below. Examination of Fig. \ref{m2l} also shows that the Millennium
Simulation prediction for the locus of mass-to-light ratio against
mass relation for data using dynamical masses and with selection
criteria applied (dashed line) is an extremely good match to
the observational data. What is more, in Table \ref{spreads} we show
the width of the 15th and 85th percentiles in dynamical mass at
selected points along the mass-luminosity relations of the simulated
and real data, and agreement is again excellent. Clearly, the
Millennium Simulation is doing a remarkably good job of reproducing the
actual observational data.

Our analysis of the Millennium Simulation has therefore clearly identified
the importance of selection effects in shaping the locus of the
mass-luminosity relation found from our data, and suggests that such
effects may be present to, some lesser extent, in the other studies
considered here. Indeed, the excellent agreement found between our
data and the Millennium Simulation suggest that the simulation might be used
to identify, and even correct for, such effects in future studies.

\section{Conclusions}
As the first step in a program to perform a detailed study of ``fossil
groups", we have estimated masses and total luminosities in $\sim$600
groups and clusters identified in the maxBCG catalogue using SDSS
photometry and spectroscopy. We have also performed an analysis of
Millennium Simulation data subjected to the same processes and
selection criteria as the observational data. We have compared our
results with five other studies in the literature in four different
wavebands ($B$, $r$, $i$ and $K$) and the Millennium Simulation data. 

At high masses ($>$10$^{14}$  M$_{\odot}$) we find extremely good agreement with
four of the five literature and the Millennium Simulation results, but
find an exception in a single literature study \citep{Gir02}, which
appears discordant. At lower masses the agreement of our data with
literature results is poorer. However, we have used the Millennium
Simulation to demonstrate that a large part of the discrepancies at
these masses can be explained by selection effects.

We have also used the Millennium Simulation to provide a hint to the
cause of remaining discrepancies at lower mass, showing that within
the simulation the assumption that mass is proportional to $\sigma^3$
begins to fail as one approaches lower masses. With hints of a similar
phenomena already having been observed in X-ray data
\citep[e.g.][]{Helsdon00, OP04}, this is an issue that certainly
warrants further investigation. It should be noted that our findings
with regard the Millennium Simulation are of particular importance in
studies that attempt to measure the gradient of the mass-to-light
ratio against mass scaling relation which, as is evident from our
results, are extremely susceptible to these effects.

\noindent{\bf Acknowledgments}\\ 

The authors would like to thank Eduardo Cypriano for many useful 
discussions.  CMdO acknowledges support from FAPESP (grant number 
2006/56213-9) and CNPq. LA acknowledges support from a CNPq PIBIC 
fellowship. The Millennium
Simulation databases used in this paper and the web application
providing online access to them were constructed as part of the
activities of the German Astrophysical Virtual Observatory
(GAVO). Funding for SDSS-III has been provided by the Alfred P. Sloan
Foundation, the Participating Institutions, the National Science
Foundation, and the U.S. Department of Energy Office of Science. The
SDSS-III web site is http://www.sdss3.org/.  SDSS-III is managed by
the Astrophysical Research Consortium for the Participating
Institutions of the SDSS-III Collaboration including the University of
Arizona, the Brazilian Participation Group, Brookhaven National
Laboratory, Carnegie Mellon University, University of Florida, the
French Participation Group, the German Participation Group, Harvard
University, the Instituto de Astrofisica de Canarias, the Michigan
State/Notre Dame/JINA Participation Group, Johns Hopkins University,
Lawrence Berkeley National Laboratory, Max Planck Institute for
Astrophysics, Max Planck Institute for Extraterrestrial Physics, New
Mexico State University, New York University, Ohio State University,
Pennsylvania State University, University of Portsmouth, Princeton
University, the Spanish Participation Group, University of Tokyo,
University of Utah, Vanderbilt University, University of Virginia,
University of Washington, and Yale University.

\label{lastpage}

\end{document}